\documentclass{iopart} 
\usepackage{iopams}
\advance \textwidth 15mm
\advance \textheight 15mm

\begin{document}
% Journal identifier can be put here if required, e.g.
\jl{03}
 
\newcommand{\btimes}{\boldsymbol{\times}}

\title[2D charged particles in a magnetic filed]%
{Algebraic description of a two-dimensional system of charged particles
in an external magnetic field and periodic potential}

\author{W Florek\footnote{Electronic mail: florek@amu.edu.pl}}

\address{Computational Physics Division, Institute of Physics, Adam
Mickiewicz University, Umultowska~85, 61--614
Pozna\'n, Poland}

\begin{abstract}
 Properties of the magnetic translation operators for a charged particle 
moving in a crystalline potential and a uniform magnetic field show that it
is necessary to consider {\it all} inequivalent irreducible projective 
representations of the the crystal lattice translation group. These 
considerations lead to the concept of magnetic cells and indicate
the periodicity of physical properties with respect to the charge.
It is also proven that a direct product of such representations describe
a system of two (many, in general) particles. Therefore, they can be 
applied in description of interacting electrons in a magnetic field,
for example in the fractional quantum Hall effect.
 \end{abstract}

%\pacs{03.65.Fd, 02.10.Sp, 71.70.Di}

%\submitted 
%\today

% \maketitle

\section{Introduction} 

 The magnetic translation operators
 \[
   T({\bi R})=\exp\left[-\frac{\rmi}{\hbar}{\bi R}\bdot
   \left({\bi p}-\frac{e}{c}{\bi A}\right)\right]
 \]
 introduced by Brown (1964), to describe the movement of a Bloch
electron in an external magnetic field, form in fact a projective (ray)
representation of the translation group with a factor system
(Brown 1964, Zak 1964a, b)
 \[
   T({\bi R})T({\bi R}')[T({\bi R}+{\bi R}')]^{-1}
    =m({\bi R},{\bi R}')=\exp[-\frac12\frac{{\rmi e}}{c\hbar}
     ({\bi R}\btimes{\bi R}')\bdot{\bi H}]
 \]
 where ${\bi H}=\bnabla\btimes{\bi A}$. This is only one of many applications
of projective representations, firstly investigated by Schur 
(1904, 1097, 1911), in quantum physics. However, its clarity and 
importance led Backhouse and Bradley to start their series of 
articles on projective representations with this example (Backhouse and 
Bradley 1970, Backhouse 1970, 1971, Backhouse and Bradley 1972). Another
important application is illustrated by the construction of space groups
(Altmann 1977); however in this case one considers projective 
representations of the point group (see also Bradley
and Cracknell 1972).

The other---equivalent---description of Bloch electrons in a magnetic 
field was proposed by Zak (1964a, b) and applied, e.g., by Divakaran
and Rajagopal (1995) and the author (Florek 1994, 1996a, b). This approach
consists in introduction of a covering group and investigations of 
its ordinary, i.e. vector, representations (see also Altmann 1977, 1986).
The covering group contains pairs $(\alpha,{\bi R})$, 
$\alpha\in {\mathrm U(1)}$, 
and its vector representation can be written as a product 
$\Gamma(\alpha)T({\bi R})$, where $\Gamma$ is a representation of 
$\mathrm{U(1)}$ and
$T$ is a projective representation of the translation group (Zak 1964a, 
Altmann 1977, Florek 1994). Zak rejected representations with 
$\Gamma(\alpha)\neq\alpha$ as `non-physical' (Zak 1964b). However, if $T'$
is a projective representation with a factor system 
$\Gamma(m({\bi R},{\bi R}'))$, then the product $\Gamma T'$ is a vector 
representation of the covering group and there are no rules that are 
contravened by considering
this case. The first attempt to consider {\it all} representations was
performed within Zak's approach by the author (Florek 1997a); in that 
work, the physical consequences of taking into account all cases were 
indicated.

This paper is based on Brown's approach; i.e. projective representations
of the translation group are considered. It is shown that all projective
representations are necessary
in a description of the movement of a particle with the charge $qe$, 
where $q$ is an integer, in a magnetic field and a crystalline potential.
Moreover, applying results of earlier articles (Florek 1997b,
Florek and Wa{\l}cerz 1998), this is done for any vector potential $\bi A$
(strictly speaking, for $\bi A$ a linear function of the 
coordinates; however, by appropriate gauge transformation each vector
potential can be written in such a form for a constant, uniform magnetic
field). This removes the restriction imposed by Brown (1964) and 
Zak (1964a, b) on ${\bi A}$ of being a fully antisymmetric function
of coordinates 
(i.e. $\partial A_l/\partial x_k+\partial A_k/\partial x_l=0$ for 
each pair $k,l=1,2,3$). Moreover, the proposed approach yields in a natural
way the concept of magnetic cells (Zak 1964a, b) and proves the periodicity
of physical properties with respect to the charge, in addition to
the periodicity with respect to the magnitude of the magnetic field proven
by Azbel (1963). Since projective representations corresponds to energy 
levels of one-particle states, their direct products must 
describe two-particle states (or many-particle states in a more general
case). A system of two particles with the charges $qe$ and $q'e$ 
has the total charge $(q+q')e$ and, therefore, should 
correspond to a projective representations with a factor system determined
by this charge. It follows from the previous discussion that 
in a many-body problem one has to 
consider all representations, also those considered by Zak as 
`non-physical'.

\section{Periodicity with respect to charge}\label{Pch}

The Hamiltonian describing the motion of a charged particle in a periodic 
potential $V({\bi r})$ and an external magnetic field 
${\bi H}=\bnabla\btimes {\bi A}$ is given as 
  \[
  {\cal H}=\frac{1}{2m}\left({\bi p}-\frac{qe}{c}{\bi A}\right)^2
  +V({\bi r})
  \]
 where $m$ denotes the effective particle mass, ${\bi p}$ its kinetic 
momentum, and $qe$, with $q\in{\mathbb Z}$ and $e>0$, its charge.
If the vector potential ${\bi A}$ is a linear function of the coordinates,
i.e. 
\[
A_\alpha=\sum_\beta a_{\alpha\beta}\beta\qquad \alpha,\beta=x,y,z
\]
 then
the magnetic translation operators can
be written as (Florek 1997b, Florek and Wa{\l}cerz 1998)
 \[
  T({\bi R})=\exp\left[-\frac{\rmi}{\hbar}{\bi R}\bdot
  \left({\bi p}-\frac{qe}{c}{\bi A}'\right)\right]
 \]
 where ${\bi A}'$ is a vector potential associated with ${\bi A}$, defined  
as 
\[
A'_\alpha=\sum_\beta a_{\beta\alpha}\beta.
\]
  It is well known (Brown 1964, 
Zak 1964a, b) that the periodic boundary conditions allow us to consider 
a two-dimensional crystal lattice (in the $xy$-plane, say) and 
${\bi H}=[0,0,H]$ perpendicular to it. Hence, any lattice
vector can be considered as two-dimensional:  
\[
{\bi R}=n_1{\bi a}_1+n_2 {\bi a}_2.  
\]
 The magnetically periodic boundary conditions (Brown 1964) yield quantization
of a magnetic flux: 
 \[
  {\bi H}\bdot({\bi a}_1\btimes{\bi a}_2)=
  \frac{2\pi}{N}\,\frac{\hbar c}{e}\frac{L}{q}
 \]
 where an integer $L$ is mutually prime with the crystal period $N$. 
Replacing the left-hand side by the flux per the unit cell
\[
\phi=(e/hc){\bi H}\bdot({\bi a}_1\btimes{\bi a}_2)
\]
 one obtains
  \begin{equation}\label{condq}
  N\phi= \frac{L}{q}.
  \end{equation}

The factor system $m({\bi R},{\bi R}')$ depends
on ${\bi A}$: for example the antisymmetric gauge 
$\case12({\bi H}\btimes {\bi r})$ gives
(Brown 1964, Zak 1964a ,b)
 \begin{equation}\label{bzfac}
  m({\bi R},{\bi R}')=\omega_N^{(1/2)L(n_2n'_1-n_1n'_2)}
  =\omega_{2N}^{L(n_2n'_1-n_1n'_2)}
  \end{equation}
 whereas for the Landau gauge ${\bi A}=[0,Hx,0]$ (and 
${\bi A}'=[-Hy,0,0]$)
  \begin{equation}\label{Lfac}
  m_N^{(L)}({\bi R},{\bi R'})=\omega_N^{L n_2n'_1}.
  \end{equation}
 In both formulae, $\omega_N=\exp(2\pi\rmi/N)$. However, the group-theoretical 
commutator is gauge-independent, and for any linear gauge we have 
(Florek and Wa{\l}cerz 1998)
 \begin{equation}\label{comm}
  T({\bi R})T({\bi R}')T^{-1}({\bi R})T^{-1}({\bi R}')=
  \omega_N^{-L (n_1n'_2-n_2n'_1)}.
  \end{equation}
 The matrices of irreducible projective representation corresponding to the 
factor system given as \eref{Lfac} can be chosen as 
  \begin{equation}\label{Lrep}
  D_{ij}^{NL}({\bi R})
  =\delta_{i,j-n_2}\omega_N^{Ln_1i}\qquad i,j=0,1,\dots,N-1\,.
  \end{equation}
 It should be underlined that such a projective representation is normalized
(cf. Altmann 1977, 1986, Florek and Wa{\l}cerz 1998), in contrast to those 
corresponding to the factor system \eref{bzfac} and considered by Brown (1964).
 
If $\gcd(L,N)=\nu>1$ the representations \eref{Lrep} are reducible and the 
corresponding factor system is
 \begin{equation}\label{fnl}
 m_n^{(l)}({\bi R},{\bi R}')=\omega_n^{l n_2n'_1}
 \end{equation}
 where $l=L/\nu$, $n=N/\nu$ and $\gcd(l,n)=1$.
 Irreducible projective representations with such factors have to be 
$n$-dimensional, which directly leads to the concept of {\it magnetic cells}:
one obtains $D^{NL}(n{\bi R})={\mathbf 1}$, so the {\it magnetic period}
is equal to $n$, though the {\it crystal period} is still $N$. Therefore,
the $N\times N$ lattice can be viewed as a $\nu\times\nu$ lattice, with the 
translation group $T_\nu= {\mathbb Z}_\nu^2$, of $n\times n$ magnetic 
cells. Let $(\xi_1,\xi_2)$ label magnetic cells, whereas 
$(\eta_1,\eta_2)$ is the position within a magnetic cell, i.e. 
$n_i=\eta_i+\xi_in$. Then matrices
 \begin{equation}\label{Dnlq}
 D_{ij}^{nl,{\bi k}}({\bi R})
 =D_{ij}^{nl}(\eta_1,\eta_2)D^{\bi k}(\xi_1,\xi_2)
 =\delta_{i,j-\eta_2}\omega_n^{l\eta_1i}D^{\bi k}(\xi_1,\xi_2)
 \end{equation}
 form an irreducible projective representation of ${\mathbb Z}_N^2$ with
the factor system \eref{fnl}, 
where
 \begin{equation}\label{Tnuk}
 D^{\bi k}(\xi_1,\xi_2)
 =\exp[-2\pi\rmi(k_1\xi_1+k_2\xi_2)/\nu]
 =\omega_{\nu}^{-(k_1\xi_1+k_2\xi_2)}
 \end{equation}
 is an irreducible representation of $T_\nu$ (Backhouse 1970).
The character of the representation given by \eref{Dnlq} is  
 \[
 \chi_{n,l;{\bi k}}({\bi R})
 =\delta_{\eta_1,0}\delta_{\eta_2,0}n \omega_{\nu}^{-(k_1\xi_1+k_2\xi_2)}.
 \]
 For given $n$ and $l$ (i.e. for a given factor system), we obtain all
$\nu^2$ inequivalent irreducible projective representations labelled
by ${\bi k}$ (Altmann 1977, 1986), and all of them are normalized.

To determine a relation between the charge $q$ of a particle and the 
irreducible projective representation $D^{nl,{\bi k}}$, let
us fix the magnetic flux $\phi$ and the crystal period $N$. Then the
condition \eref{condq} gives that $L=N\phi q$; i.e. $L\propto q$. 
However, this is not
a one-to-one relation, since $L$ is limited to the range $0,1,\dots,N-1$ with
no condition imposed on $q\in{\mathbb Z}$. The representation 
\eref{Lrep}, its factor system \eref{Lfac}, and the commutator \eref{comm}
are determined by $\omega_N^L$, so all of them are periodic functions
of $L\propto q$, and, therefore, periodic functions with respect to
the charge of a moving particle. We see, in particular, that for 
$q=zN$, $z\in{\mathbb Z}$, vector representations with trivial factor
systems (and trivial commutators) are obtained. This means that for a 
given crystal period $N$ and constant magnetic field, a particle with the 
charge $zNe$ behaves as non-charged one. It is also easy to see that 
for some $q$ we can obtain $L=l\nu$, where $\nu=\gcd(N,L)$, and in this 
case the irreducible representations $D^{nl,{\bi k}}$ have to be used. 
Since $\nu$ is a co-divisor of $n$, then assuming $\phi=1/N$ we obtain
 \begin{equation}\label{nlq}
 q=N\frac{l}{n}
 \end{equation}
 which relates the pair $(n,l)$ (the label of the irreducible representation)
and the charge $q$ of a particle. It has to be underlined that this 
relation has been derived for a fixed $\phi$ and 
does not depend on the irreducible 
representations $D^{\bi k}$ of $T_\nu$ given by \eref{Tnuk}.

\section{Multi-particle states}\label{Pipr}

It can be shown (see, for example, Altmann 1986) that a product of two 
projective 
representations $D'$ and $D''$ of a given group $G$ with factor systems $m'$ 
and $m''$, respectively,  is another projective representation
with a factor system $m(g,g')=m'(g,g')m''(g,g')$, which, in general, is 
different from factor systems $m'$ and $m''$.
Let $D$ be a product of two irreducible projective representations 
$D^{nl,{\bi k}}$ and $D^{n' l',{\bi k}'}$. Then their product has
a factor system 
 \begin{equation}\label{fprod}
 m({\bf R},{\bf R}')=\omega_N^{Ln_2n'_1}\qquad \mathrm{with}\;
 L=l\nu+l'\nu'
 \end{equation}
 so it corresponds to the representation $D^{NL,{\bi K}}$ (${\bi K}$
has not been determined, but it depends on the irreducibility of
the representation obtained). The character of this representation is
 \[
 \chi({\bi R})
 =\delta_{\eta_1,0}\delta_{\eta_2,0}\delta_{\eta'_1,0}
 \delta_{\eta'_2,0}\,nn'
 \omega_N^{-n(k_1\xi_1+k_2\xi_2)-n'(k'_1\xi'_1+k'_2\xi'_2)}
 \]
 so it is nonzero only for $n_i=x_i m$, where $m=nn'/\gamma$,
$\gamma=\gcd(n,n')$, 
$0\le x_i<\mu=N/m=\gcd(\nu,\nu')$. Substituting
$m$ and $\mu$ to the above formula one obtains
 \begin{equation}\label{chiD}
 \chi({\bi R})
 =\delta_{\eta_1,0}\delta_{\eta_2,0} m\gamma
 \omega_{\mu}^{-(k_1+k'_1)x_1-(k_2+k'_2)x_2}\pmod m.
 \end{equation}
 Since $\nu/\mu=n'/\gamma$, then $L$ in \eref{fprod} can be written as
 \begin{equation}\label{fprm}
 L=\mu\left(\frac{l\nu}{\mu}+\frac{l'\nu'}{\mu}\right)
 =\mu\left(\frac{ln'}{\gamma}+\frac{l'n}{\gamma}\right)
 =\mu\lambda.
 \end{equation}
 It seems that this determines a factor system $m_m^{(\lambda)}$.
However, we cannot exclude the case in which
$\gcd(\lambda,m)=\ell>1$. Therefore, the product
considered has to be decomposed into irreducible representations with
a factor system $m_M^{(\Lambda)}$, where $\Lambda=\lambda/\ell$ and
$M=m/\ell$. The scalar product of the appropriate characters gives us
a multiplicity of $D^{M\Lambda,{\bi K}}$ in the product considered, as
follows:
\begin{equation}
f(D^{M,\Lambda;{\bi K}}, D^{nl,{\bi k}}\otimes D^{n'l',{\bi k}'})
=\frac{\gamma}{\ell}\delta_{K_1,k_1+k'_1}
 \delta_{K_2,k_2+k'_2}.
 \label{freq}
 \end{equation}
There are $\ell^2$ such representations with $K_i=(k_i+k'_i) \bmod
\mu$.

The most interesting is the case when $n=n'$ and $l=l'$, since $n$
and $l$ are determined by the magnetic flux, the charge, and the
crystal period $N$; hence such a case can be interpreted as a system of two
identical particles moving in the same lattice and the same magnetic field
(Florek 1997a). The resultant representation is
$n^2$-dimensional and its character is equal to
 \[
 \chi({\bi R})
 =\delta_{n_1,x_1n}\delta_{n_2,x_2n} n^2
 \omega_{\nu}^{-(k_1+k'_1)x_1-(k_2+k'_2)x_2}
 \]
 with $0\le k_i,k'_i,x_i<\nu$. The factor system is given by
\eref{fprod} as
 \[
 m({\bi R},{\bi R}')=\omega_n^{2ln_2n'_1}
 =\omega_n^{\lambda n_2n'_1}
 \]
 so we have to check the $\gcd(\lambda,n)$. At this moment the cases of
odd and even $n$ have to be considered separately. In the first case,
$\ell=\gcd(n,2l)=1$ and the representations obtained decomposes into
$n$ copies of the representation $D^{n2l,{\bi K}}$ with
$K_i=(k_i+k'_i)\bmod \nu$. In the second case, however, $\ell=2$ and
$M=\case12n$, so the considered product decomposes into
representations $D^{\case12nl,{\bi K}}$: there are four
representations with $K_i=(k_i+k'_i)\bmod \nu$ and each of them appears
$\case12n$ times. In both cases we have 
 \[\frac{2l}{n}=\frac{l}{(n/2)}=2\frac{l}{n}\]
 so the new representations correspond to a system with the charge $2q$, see
\eref{nlq}. 
However, an even $n$ in the second case yields the change of magnetic 
periodicity from $n$ to $\case12n$ and four times as many magnetic cells. 
In a similar way, the coupling 
of $d$ representations $D^{n1,{\bi k}^{(j)}}$, $j=1,2,\dots,d$
with $n=dM$, changes the magnetic period from $n$ to $M$ (and yields 
$d^2$ times as many magnetic cells)---however, not
by modification of the magnetic field, but by multiplication of the
charge by $d$. 

The irreducible representations \eref{Dnlq} are written as 
a product of a one-dimensional irreducible representation 
$D^{\bi k}$ of $T_\nu$, equation
\eref{Tnuk}, and a projective one of $T_n$. It means that also products of 
such representations can also be separated into a part describing addition of 
the quasi-momenta ${\bi k}$, ${\bi k}'$ 
with the second part corresponding to the addition of co-divisors $\nu$ and 
$\nu'$ or, more precisely, $l\nu+l\nu'$, see \eref{fprod}. However, 
the last addition can change the magnetic periodicity, determined by 
$M$ and $\Lambda$ in \eref{fprm} and \eref{freq}, in a way depending on 
the arithmetic structure of $N$, $n$, $n'$, $l$, and $l'$. In the above 
example, the label $M$ (the size of magnetic cells) of 
resultant representation was equal to or smaller than $n=n'$. 
One can easily obtain that for $N=12$
 \[
 D^{3,1;[1,0]}\otimes D^{6,1;[1,0]}=
 \bigoplus_{K_1,K_2=0,2,4}
 D^{2,1;[K_1,K_2]}.
 \]
 In this case one particle may have charge $4e$ and the second $2e$, so
the two-particle system has the charge $6e$. We must say `may have' since
the condition \eref{condq} involves both the magnetic flux $\phi$ and the
charge $q$. The chosen values of charges correspond to the fixed 
$\phi=1/N$. Therefore, the charge of the first particle
yields $3\times3$ magnetic cells, and the second one $6\times6$, whereas
two-particle system demands $2\times2$ magnetic cells. On the other hand
we have ($N=12$, as above)
 \[
 D^{3,1;[1,0]}\otimes D^{4,1;[1,0]}=
 D^{12,7;[0,0]}
 \]
 so $M>n,n'$ and there is only one magnetic cell. Therefore, the addition 
of quasi-momenta ${\bi k}$, ${\bi k}'$ has to be modified to reflect all 
possible changes of the magnetic periodicity.

\section{Final remarks and conclusions}

The projective representations used by Brown (1964) and in this paper
can be replaced in an equivalent approach by using vector representations
of central extensions (Zak 1964a, b, Florek 1994, 1996a, b). Zak assumed
that a factor $\omega$ has to be represented by itself, and rejected
representations in which $\omega$ is represented by $\omega^r$. However,
as long as $r$ is mutually prime with $N$, such a change is an isomorphism of
(inequivalent) central extensions (Altmann 1977,
Florek 1994). Within the approach presented here
this fact is realized by the
freedom that one has 
in choosing relation between the charge $q$ and the index $l$, given by
\eref{nlq}. For $q=1$, we can take not only $L=1$ but also any $r$ mutually
prime with $n$. All important properties, e.g. addition of charges and
charge periodicity, are unaffected: since $\gcd(r,n)=1$, then 
$\{r,2r,\dots,Nr\}=\{1,2,\dots,n\}$, but elements of the first set 
are obtained in a 
different order ($zr$ is calculated $\bmod\, n$).
In physical terms, this means that if we observe only magnetic or charge 
periodicity,
we cannot distinguish $H_1=2\pi\hbar c/Ne$ form $H_r=rH$ if $\gcd(r,N)=1$;
see \eref{condq} and \eref{nlq}. In fact, it should be said that the 
condition \eref{condq} is not imposed on $H$ or $q$ but on their product $qH$,
and has to be written as 
 \begin{equation}\label{condqH}
 qH=\frac{2\pi}{N}\,\frac{\hbar c}{e}L\quad \mathrm{or}\quad
 q\phi =\frac{L}{N}\,.
 \end{equation}
 This means that a particle with the charge $2e$ can be described by the same
representation $D^{NL}$ as a particle with the charge $e$ if the magnetic 
field is halve.  On the other hand, very strong magnetic fields
may lead to observations of a fractional charge, if the product
$qH$ has to satisfy \eref{condqH}.

The introduction of projective representations in this paper has been based 
on the magnetic translation operators determined by Brown (1964), and the 
notion of Bloch electrons in an external magnetic field was used throughout
this work. Hence, the concept of magnetic cells has appeared
in a natural way. However, these representations can be applied to any 
problem in quantum mechanics in which a symmetry group $G$ appears and 
phase factors play an important role. For example, Divakaran and Rajagopal
 (1991)
used them in the theory of superconducting layered materials (they  
included also many general remarks in their work). If we assume that projective
representations correspond to energy levels (and so representation vectors 
correspond to states) of a one-particle system, then products of two 
(or more) representations
have to correspond to two-particle (or many-particle, in a general case) 
systems.  Not straying far
from physical problems discussed above, we can look at a 
two-dimensional electron
gas in an external magnetic field. The fractional quantum Hall effect (Tsui 
\etal 1982, Das Sarma and Pinczuk 1997) is still a subject to which much 
effort is being devoted by 
theorists and experimentalists, but it has been accepted that Coulomb
interactions play a very important role in explanation of observed features
(Shankar and Murthy 1997, Heinonen 1998). Therefore, it seems 
possible to apply the results presented above to such problems. 

It should be underlined that products of {\it projective} representations are
well known in mathematics (Backhouse and Bradley 1972, Altmann 1986).
On the other hand, products of {\it vector} representations are commonly
used in quantum physics to describe multi-particle states. 
It is shown in this paper that {\it products of projective} 
representations also have to be applied in many-body problems. 

\ack

It is a pleasure to thank Professor G.~Kamieniarz for carefully reading
the manuscript and many helpful remarks. Partial support from the State
Committee for Scientific Research (KBN) within
 the project No 8~T11F~027~16 is acknowledged.

\section*{References}

 \begin{harvard}
 \item[] Altmann S L 1977 {\it Induced Representations in Crystals and
Molecules} (London: Academic Press) 
 \item[] \dash 1986 {\it Rotations, Quaternions, and Double Groups}
 (Oxford: Clarendon)
 \item[] Azbel M Y 1963 {\it Zh. Eksp. Teor. Fiz.} {\bf44} 980 
(Engl. Transl. 1963 {\it Sov. Phys. JETP} {\bf 17} 665)
 \item[] Backhouse N B 1970 {\it Quart. J. Math. Oxford} {\bf 21} 277
 \item[] \dash 1971 {\it Quart. J. Math. Oxford} {\bf 22} 277 
 \item[] Backhouse N B and  Bradley C J 1970 
{\it Quart. J. Math. Oxford} {\bf 21} 203
 \item[] \dash 1972 {\it Quart. J. Math. Oxford} {\bf 23} 225 
 \item[] Bradley C J and Cracknell A P 1972 {\it The Mathematical Theory
  of Symmetry in Solids} (Oxford: Clarendon)
 \item[] Brown E 1964 \PR {\bf 133} A1038
 \item[] Das Sarma S and Pinczuk A {\it Perspectives in the Quantum Hall Effect}
 (New York: Wiley) 
 \item[] Divakaran P P and Rajagopal A K 1991 {\it Physica} C 
{\bf 176} 457
 \item[] \dash 1995 {\it Int. J. Mod. Phys.} B 
{\bf 9} 261
 \item[] Florek W 1994 {\it Rep. Math. Phys.} {\bf 34} 81
 \item[] \dash 1996a {\it Rep. Math. Phys.} {\bf 38} 235
 \item[] \dash 1996b {\it Rep. Math. Phys.} {\bf 38} 325
 \item[] \dash 1997a \PR B {\bf 55} 1449 
 \item[] \dash 1997b {\it Acta Phys. Pol.} A {\bf 92} 399
 \item[] Florek W and Wa{\l}cerz S 1998 \JMP {\bf 39} 739
 \item[] Heinonen O 1998 {\it Composite Fermions} (New York: World Scientific)
 \item[] Schur I 1904 {\it J. reine angew. Math.} {\bf 127} 20
 \item[] \dash 1907 {\it J. reine angew. Math.} {\bf 132} 85
 \item[] \dash 1911 {\it Math. Ann.} {\bf 71} 85 
 \item[] Shankar R and Murthy G 1997 \PRL {\bf 79} 4437
 \item[] Tsui D, St\"ormer H L and Gossard A C  1982 \PRL {\bf 48} 1559
 \item[] Zak J 1964a \PR {\bf 134} A1602
 \item[] \dash 1964b \PR {\bf 134} A1607

\end{harvard}

\end{document}